\documentclass[10pt,aps,prb,twocolumn,showpacs,amsmath,amssymb,superscriptaddress]{revtex4-1}
\usepackage{graphicx}
\usepackage{dcolumn}
\usepackage{bm}
\usepackage{epsfig}
\usepackage{multirow}
\usepackage{rotating}
\usepackage{xcolor}

\begin{document}
\title{Spin diffusion in the low-dimensional molecular quantum Heisenberg
  antiferromagnet Cu(pyz)(NO$_{3}$)$_{2}$ detected with implanted muons}
\author{F. Xiao}
\affiliation{Durham University, Department of Physics, South Road,
  Durham, DH1 3LE, UK}
\author{J.S. M\"{o}ller}
\altaffiliation{Present address: Neutron Scattering and Magnetism, Laboratory for Solid State Physics, ETH Z\"{u}rich, CH-8093 Z\"{u}rich, Switzerland}
\affiliation{University of Oxford, Department of Physics, Clarendon
  Laboratory, Parks Road, Oxford, OX1 3PU, UK}
\author{T. Lancaster}
\affiliation{Durham University, Department of Physics, South Road,
  Durham, DH1 3LE, UK}
\author{R.C. Williams}
\affiliation{Durham University, Department of Physics, South Road,
  Durham, DH1 3LE, UK}
\author{F.L. Pratt}
\affiliation{ISIS Pulsed Neutron and Muon Facility, STFC Rutherford
  Appleton Laboratory, Harwell Oxford, Didcot, OX11 OQX, UK}
\author{S.J. Blundell}
\affiliation{University of Oxford, Department of Physics, Clarendon
    Laboratory, Parks Road, Oxford, OX1 3PU, UK}
\author{D. Ceresoli}
\affiliation{Istituto di Scienze e Tecnologie Molecolari CNR, via Golgi 19, 20133 Milano, Italy}
\author{A.M. Barton}
\affiliation{Eastern Washington University, Department of Chemistry
  and Biochemistry, Cheney, Washington 99004, USA}
\author{J.L. Manson}
\affiliation{Eastern Washington University, Department of Chemistry
  and Biochemistry, Cheney, Washington 99004, USA}
\date{\today}

\begin{abstract}
We present the results of muon-spin relaxation measurements of spin excitations in the one-dimensional quantum Heisenberg antiferromagnet Cu(pyz)(NO$_{3}$)$_{2}$. Using density-functional theory we propose muon sites and assess the degree of perturbation the muon probe causes on the system. We identify a site involving the muon forming a hydroxyl-type bond with an oxygen on the nitrate group that is sensitive to the characteristic spin dynamics of the system. 
Our measurements of the spin dynamics show that in the temperature range $T_{\mathrm{N}}<T<J$ (between the ordering temperature $T_{\mathrm{N}}$ and the exchange energy scale $J$) the field-dependent muon spin relaxation is characteristic of diffusive transport of spin excitations over a wide range of applied fields. We also identify a possible crossover at higher applied fields in the muon probe's response to the fluctuation spectrum, to a regime where the muon detects early-time transport with a ballistic character. 
This behavior is contrasted with that found for $T>J$ and that in the related two-dimensional system Cu(pyz)$_2$(ClO$_4$)$_{2}$.

\end{abstract}
\maketitle
\section{Introduction}
Low-dimensional quantum magnetism continues to be of great theoretical and experimental interest as reduced dimensionality supports strong quantum fluctuations which can result in novel excitations and critical behavior~\cite{sachdev-qpt}. A notable challenge in this field is the elucidation of the mechanism for the transport of spin excitations in the one-dimensional quantum Heisenberg antiferromagnet (1DQHAF). Intuitively it might be expected that the system obeys a standard diffusion phenomenology. However, in the absence of any microscopic formalism this has been vigorously debated. 
Spin transport in this system, whose Hamiltonian is given by $\mathcal{H}=J\sum\mathbf{S}_i\cdot\mathbf{S}_{i+1}$,  has been extensively studied~\cite{sachdev-prb-1994, castella-prl-1995, takigawa-prl-1996, takigawa-prb-1997, narozhny-prb-1998, zotos-prl-1999, heidrich-meisner-prb-2003, benz-jpsjs-2005, sirker-prl-2009, langer-prb-2009, grossjohann-prb-2010,sirker-prb-2011,znidaric-prl-2011,steinigeweg-epl-2012} but the nature of transport remains controversial, with some theoretical and numerical studies~\cite{castella-prl-1995, narozhny-prb-1998, benz-jpsjs-2005} suggesting that the transport is necessarily ballistic; while others~\cite{zotos-prl-1999, grossjohann-prb-2010} show it to be diffusive. More recent work has shown that when subjected to a periodic lattice potential, diffusion can co-exisit with ballistic transport~\cite{sirker-prl-2009,sirker-prb-2011}. 

Muon spin relaxation ($\mu^+$SR) has been shown to be a sensitive probe of low-dimensional magnets, not only revealing the ordering temperatures in such a system~\cite{tom-prb-2006}, but also being sensitive to the nature of the spin excitations in the MHz frequency range, allowing the identification of diffusive and ballistic behavior. For example, the study of the molecular radical 1DQHAF DEOCC-TCNQF$_{4}$ identified diffusive transport~\cite{francis-prl-2006} while in the inorganic 1DQHAF Rb$_{4}$Cu(MoO$_{4}$)$_{3}$ ballistic transport was shown to dominate~\cite{tom-prb-2012}. 

Many coordination polymer molecular magnets have been shown to closely realize models of low-dimensional magnetism, with the advantage that their typical energy scales are experimentally accessible in contrast to many oxide materials. Coordination polymers comprise regular arrays of transition metal ions (e.g.\ $S=$ 1/2 Cu$^{2+}$) linked with molecular ligands such as pyrazine (pyz=C$_{4}$H$_{4}$N$_{2}$). Cu(pyz)(NO$_{3}$)$_{2}$ consists of well-isolated Cu-pyz-Cu chains and has been shown to be a highly successful realization of the 1DQHAF~\cite{hammar-prb-1999,kono-prl-2015}, with a principal exchange constant of $J= 10.3(1)$~K [$\equiv$ 7.2(1)~T~\footnote{The equation $g\mu_\mathrm{B}J[\mathrm{T}]=k_\mathrm{B}J[\mathrm{K}]$ was used to calculate the equivalent value of $J$ in magnetic field units. The average $g$-value (2.13) was taken from Ref~\onlinecite{hammar-prb-1999}.}]. 
Although the ground state of an ideal 1DQHAF is a Tomonaga-Luttinger liquid with a gapless spectrum of excitations with linear dispersion, a real system is expected to undergo long-range magnetic ordering due to the weak coupling between spin chains. Previous $\mu^{+}$SR measurements~\cite{tom-prb-2006} showed that Cu(pyz)(NO$_{3}$)$_{2}$ orders at $T_{\mathrm{N}}= 0.107(1)$~K, leading to an estimate of the interchain-intrachain coupling ratio of $J_\perp/J_\parallel\approx4.4\times10^{-3}$. Inelastic neutron scattering (INS) measurements at 0.25~K reveal the expected spinon excitation spectrum~\cite{hammar-prb-1999, stone-prl-2003} while NMR experiments demonstrate that a shift of the maximum of field-dependent relaxation rate $T_1^{-1}$ in the system is caused by spin-spin interactions~\cite{kuhne-prb-2011}. 

\begin{figure*}
\includegraphics[width=0.95\linewidth]{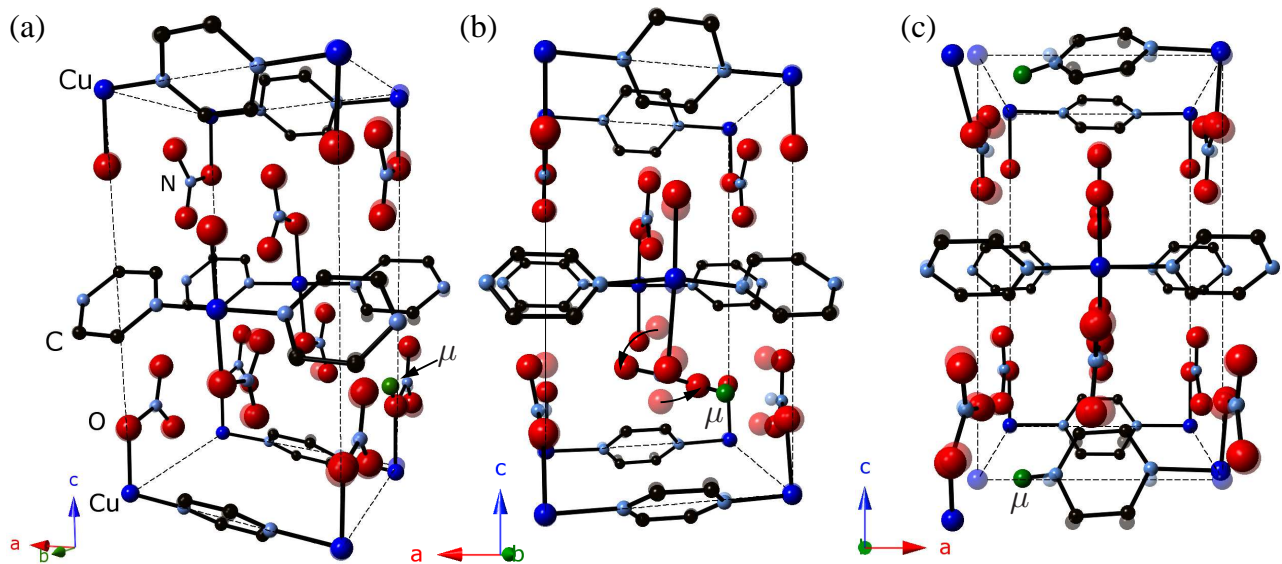}
\caption{\label{fig:structures} (Color online) The low-energy muon sites in Cu(pyz)(NO$_3$)$_2$. Translucent spheres represent the ionic positions in the unit cell without the muon. (a) The nitrate site, first configuration (bonding oxygen moved by $0.4$~\AA, other ions move by $\leq 0.2$~\AA). (b) The nitrate site, second configuration, with the nearly 90$^\circ$ rotation of the nitrate group around the $b$-axis (oxygen ions moved by 1.5, 1.2, 0.3~\AA). (c) The N(pyz) site (Cu ion moved by 1.4~\AA). }\label{fig:sites}
\end{figure*}

In this paper we present the results of $\mu^+$SR measurements of the 1DQHAF Cu(pyz)(NO$_{3}$)$_{2}$ in the temperature regime $T_{\mathrm{N}}<T<J$, where we access the excitations that result from this low-dimensional behavior. Here we show that the excitations detected by the muon are diffusive for a large range of applied magnetic fields. We contrast this behavior with the excitations probed for $T>J$. We also present results of measurements on the related 2D material Cu(pyz)$_2$(ClO$_4$)$_2$ which belongs to a family of 2DQHAFs with $J=$17.5~K and $T_\mathrm{N}=$4.2~K~\cite{woodward-ic-2007,fan-prb-2009,tom-prb-2007}. To complement our $\mu^+$SR results described below, along with those of the previous zero-field (ZF) muon measurements~\cite{tom-prb-2006} that revealed $T_{\rm N}$, we have performed density-functional theory (DFT) calculations to investigate the stopping sites of the muon in Cu(pyz)(NO$_3$)$_2$ and the extent of the perturbation the muon exhibits locally. 
DFT also allows us to determine whether some possible muon states could have level crossings in the field range used to investigate the dynamics of propagating spin excitations, which would complicate the analysis.

\section{DFT calculations and muon sites}

Recently, developments in using DFT techniques to solve the muon-site problem~\cite{johannes-scripta-2013} have been applied successfully to a variety of different systems.  These include ionic insulators~\cite{johannes-prb-2013,bernardini-prb-2013}, organic magnets~\cite{steve-prb-2013}, pnictide superconductors~\cite{prando-prb-2013} and quantum spin ices~\cite{foronda-prl-2015} but not, until now, coordination polymers. Our DFT calculations were performed with the {\sc quantum espresso} package~\cite{qespresso} within the generalized-gradient approximation~\cite{pbe} (GGA) using norm-conserving and ultra-soft\cite{uspp} pseudopotentials. The muon was modelled by a norm-conserving hydrogen pseudopotential. 
The details of the DFT calculations may be found in the Appendix.
 The results reported here were obtained in calculations for a supercell of  $2\times2\times1$ conventional unit cells (plus the muon). Both neutral and positively charged (+1) supercells were studied. The former corresponds to the scenario where the muon attracts an electron through some thermal or epithermal process as it stops in the crystal; the latter corresponds to the case when this does not occur.  

Structural relaxations of the system reveal two classes of low-energy muon sites. In the first class (denoted the NO$_{3}^{-}$ sites) the muon forms a hydroxyl-type bond with any one of the three inequivalent oxygen ions in the nitrate group.
Two configurations are possible within this class.
In the first, which occurs in both neutral and charged supercells, the muons are approximately coplanar with the nitrogen and oxygen ions (within $3^\circ$) [Fig.~\ref{fig:sites}(a)]. As we argue below, it is this site that is probing the spin dynamics in this system. We note that the perturbations caused by the muon in this site are benign and therefore do not expect our conclusions on the spin dynamics to be affected by muon-induced effects.
In the second, which occurs only in the neutral supercell,
the entire bonding nitrate group  rotates by nearly 90 degrees around the $b$-axis [Fig.~\ref{fig:sites}(b)]. 
For these sites, the crystallographic distortions are $\leq$ 0.6~\AA~for those atoms not in the bonding nitrate group. 
In the second class of muon sites [denoted N(pyz)] the muon bonds to one of the two equivalent nitrogen atoms in the pyrazine ring  [Fig.~\ref{fig:sites}(c)]. This site occurs for both neutral and charged supercells.
In this class, the nitrogen atom with the attached muon is very close to the Cu ion and this causes the crystallographic distortion of the Cu ion to be significant, involving distortions of $>1$~\AA. 

Given these proposed sites, we may assess the possible influence of the muon on the local magnetism of the system via the calculated spin density (see Appendix). For the cases involving a neutral supercell, where the muon and an extra electron are added to the system, the muon in both the NO$_3^-$ and N(pyz) sites donates the extra electron to the nearest-neighbor Cu ion, turning the magnetic Cu$^{2+}$ ion into diamagnetic Cu$^{+}$. This acts to interrupt the Cu--pyz--Cu exchange pathways, effectively cutting the spin chain. 
Furthermore, for the N(pyz) site, the Cu$^{+}$ environment distorts away from square-planar towards a linear N--Cu--N ``dumbbell'' arrangement, consistent with the observation that Cu$^{+}$ $3d^{10}$ ions prefer linear coordination in order to reduce the orbital overlap with the ligands. 
This results in the combined effect of (i) switching off the Cu moment and (ii) significantly displacing the Cu ion. 
For the charged supercell, where there is no extra electron introduced, only the structural distortion occurs. For the charged supercell N(pyz) site, for example, where only the Cu ion is displaced, this leads to an increased magnetic overlap between neighbouring Cu--pyz--Cu chains. The charged cell NO$_{3}^{-}$ site involves a less significant structural distortion still and so should be expected to have the least effect on the magnetic properties. 

In addition to our DFT results, we also note that possible states formed by a single spin $S=1/2$ impurity coupling to a spin-chain have been previously studied in detail theoretically, both using conformal field theory and numerically~\cite{eggert-prb-1992}. The result is a prediction that the coupling of a muon impurity to a $S=1/2$ Cu$^{2+}$ spin in a 1DQHAF would result in a screened spin with effective $S=0$, whose influence on the local susceptibility is healed with increasing distance from the defect. There is some evidence that such states may be realized in 1DQHAFs~\cite{chakhalian-prl-2003}.

Given these results, one might legitimately worry that if the neutral supercell sites, or the site investigated via field theory, are realized then the muon will not be an `innocent' probe of the static magnetism and dynamics of the Cu(pyz)(NO$_3$)$_2$ system. However, the results of $\mu^{+}$SR measurements on this system demonstrate that this is not the case and that the muon
may be used to investigate the intrinsic magnetic properties of the material. 
Previously~\cite{tom-prb-2006}, $\mu^{+}$SR measurements showed a crossover to a regime of static long-range magnetic order at low temperature. This was observed via well-resolved oscillations in the muon-spin relaxation spectra, occurring at two distinct frequencies in the magnetically ordered state. In addition, there was evidence for one other set of muon sites which did not give rise to a resolvable precession signal. Since the observed precession frequencies are of the expected magnitude and show standard critical behaviour around the same transition temperature as subsequently observed in specific heat studies, we may infer that the most significant part of the resolved signal results from muon sites which are sensitive to the intrinsic magnetism of the system and are highly unlikely to involve a change in charge state of the nearest neighbour copper. It is therefore probable that the two precession frequencies result from nitrate sites realized in the charged cell, where neither the magnetic nor crystal structure are significantly perturbed and where the muon is unlikely to have a strong coupling to a single Cu$^{2+}$ and enter into the singlet state.

It therefore seems likely that the muon site which is sensitive to the characteristic spin dynamics of Cu(pyz)(NO$_3$)$_2$ reported here is also one forming a hydroxyl-type bond with an oxygen on the nitrate group with no additional electron involved. 
Further supporting evidence for the occurrence of this site is presented in the discussion section below, where consistency is found between the predicted hyperfine coupling and that found on the basis of the muon measurements of the dynamics of the system. 
Some fraction of the remainder of the $\mu^{+}$SR signal (i.e.\ the non-oscillatory contribution) might then result from the N(pyz) sites and the more perturbative NO$_{3}^{-}$ site, whose interaction with the spin chain would be more complicated. 
However, even if the remaining sites do include those which strongly perturb the system and cause singlets, they will not affect the signal from those sites that are sensitive to the intrinsic magnetism and its dynamics. It is important to note that in a $\mu^{+}$SR experiment, muons are implanted in the ultradilute limit, so that the chance of muons experiencing the effect of perturbations
induced by other muons is effectively zero. Moreover, it is highly unlikely that these perturbations would be capable of significantly altering the collective magnetic behavior of the material, by condensing long-range magnetic order, for example, or by interrupting the spin transport that we investigate below. 



Finally, to motivate the interpretation of our measurements described below, 
we may predict which (avoided) level crossing resonances are expected on the application of an external magnetic field.  We only consider  $\Delta M=1$ transitions, as they are broad and dominant in the solid state, and thus could interfere most with the overall field dependence. There are two sites where the muon has a sizeable hyperfine contact coupling (see Appendix): one of the nitrate sites in the cell  (38~MHz) and the carbon sites ($\geq$370~MHz). 
The $\Delta M=1$ resonances for muon and carbon are at  $\approx$0.14~T and 1.37~T (for $A=370$ MHz), respectively. The former is within the range of fields studied. However, level crossing resonances of molecular radicals are typically quite narrow (typically 10s of mT) and hence do not affect the conclusion drawn from the scaling behaviour of $\lambda$ on much broader scales in field. 

\section{$\mu^+$SR results and Discussion}

\begin{figure}[h]
\centering
\includegraphics[width=\linewidth]{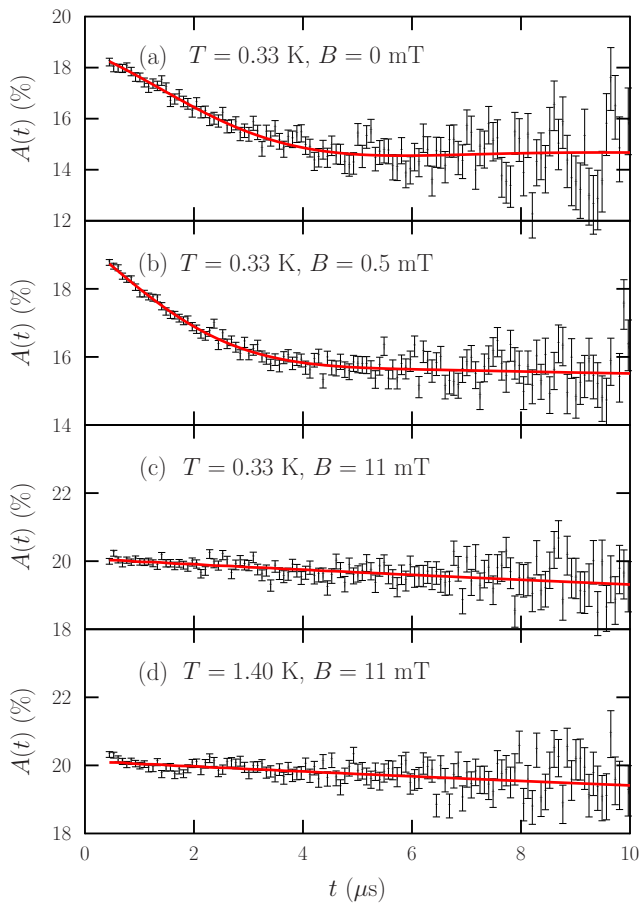}
\caption{\label{fig:muon-spectra} (Color online.) Example $A(t)$ spectra for Cu(pyz)(NO$_3$)$_2$ at (a) $T=0.33$~K and $B=0$~mT (b) $T=0.33$~K and $B=0.5$~mT (c) $T=0.33$~K and $B=11$~mT (d) $T=1.40$~K and $B=11$~mT. Solid lines represent the fits described in the text.}
\end{figure}
\begin{figure*}[t]
\centering
\includegraphics[width=\linewidth]{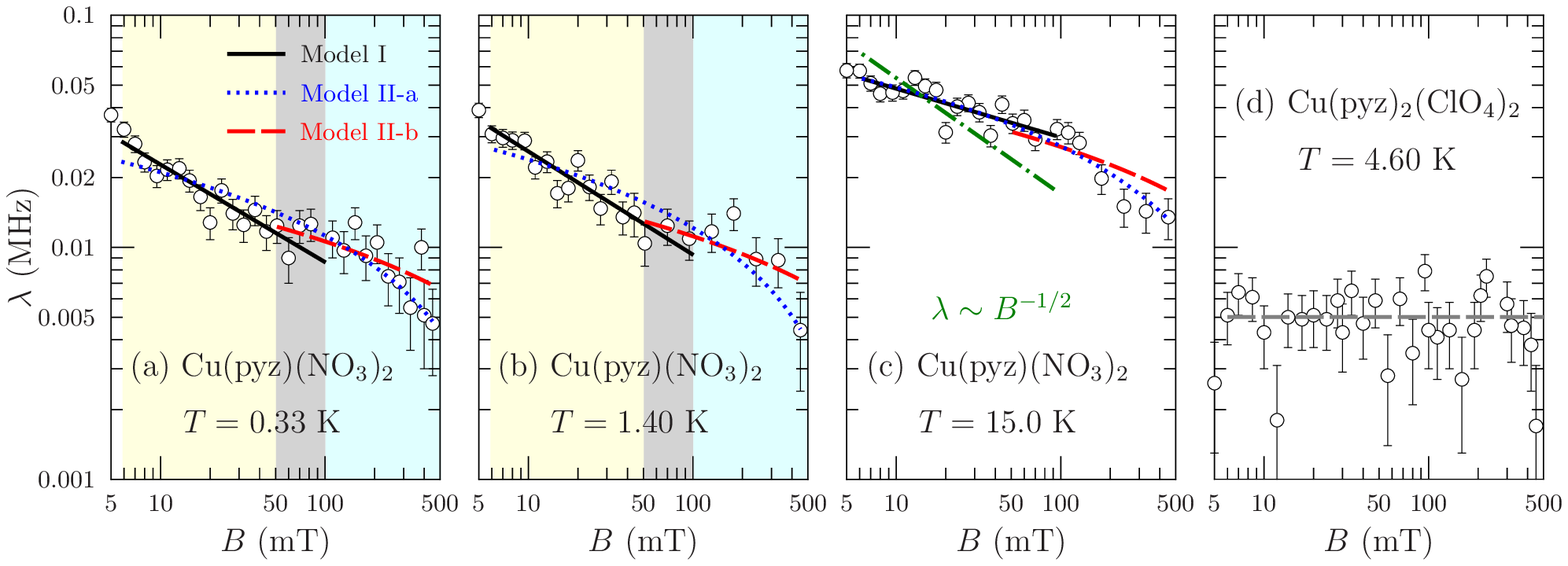}
\caption{\label{fig:fit-parameters} (Color online.) (a-c) Relaxation rates $\lambda$ for Cu(pyz)(NO$_{3}$)$_{2}$. Solid lines: power law fit; dotted lines: global fit to ballistic behavior; dashed lines: fit to ballistic behavior between 50~mT and 500~mT with $J$ fixed at 7.2~T. (c) dot dash line: an example power-law function with fixed $n= 0.5$. (d) Relaxation rates $\lambda$  for Cu(pyz)$_2$(ClO$_4$)$_2$; dashed line: linear fit function.}
\end{figure*}
Having investigated the muon sites and possible resonances expected for this material, we now turn to the results of our
investigations into the spin dynamics in the system. 
Longitudinal field (LF)  $\mu^{+}$SR measurements were made on powder samples of Cu(pyz)(NO$_{3}$)$_{2}$ and Cu(pyz)$_2$(ClO$_4$)$_2$ using the EMU spectrometer at the ISIS facility, UK at several temperatures [$T=0.33$~K, 1.4~K and 15~K for Cu(pyz)(NO$_{3}$)$_{2}$ and $T=4.6$~K for Cu(pyz)$_2$(ClO$_4$)$_2$]. 
Samples were packed in Ag foil envelopes (foil thickness 12.5~$\mu$m) and mounted on an Ag plate using vacuum grease before being attached to either the cold finger of a sorption cryostat (for $0.3<T<1.5$~K) or loaded into a $^{4}$He cryostat for $T>1.5$~K measurements. The applied external magnetic field is directed parallel to the initial muon spin directions and decouples the contribution from static magnetic fields at the muon site. This allows us to probe the dynamics of the system, as time-varying magnetic fields at the muon site are able to flip muon spins.

Example LF $\mu^{+}$SR spectra measured for Cu(pyz)(NO$_{3}$)$_{2}$ at temperatures  $T>T_{\mathrm{N}}$ are shown in Fig.~\ref{fig:muon-spectra}. In applied magnetic fields of $B\lesssim0.5$~mT, the asymmetry shows Kubo-Toyabe relaxation~\cite{hayano-prb-1979} due to the disordered magnetic fields from the nuclei surrounding the muon sites. The contribution from the nuclei, however, is suppressed when $B$ increases and the asymmetry shows only a slow exponential relaxation. 
The asymmetry $A(t)$ spectra were fitted to the function
$A(t) = A_{\mathrm{rel}} G^\mathrm{KT}_z(\Delta, B){\rm e}^{-\lambda t}+A_{\mathrm{bg}},$
which includes the contribution from both the nuclei and electronic moments. Here $A_{\mathrm{rel}}$ is the amplitude of the relaxing component, $G^\mathrm{KT}_z(\Delta, B)$ is the Kubo-Toyabe relaxation function, $\lambda$ is the relaxation rate reflecting slow dynamics of the electronic spins and $A_{\mathrm{bg}}$ accounts for the constant background contribution. For both Cu(pyz)(NO$_{3}$)$_{2}$ and Cu(pyz)$_2$(ClO$_4$)$_2$ the field width $\Delta$ was fixed at around 0.3 MHz. The values of $\lambda$ from the fitting routine are plotted against applied magnetic field in Fig.~\ref{fig:fit-parameters}. In all cases described below, the relaxation rates show a similar (approximately power-law-like) decrease at the low fields ($B\lesssim6$~mT) before developing distinct behaviors at higher fields ($B\gtrsim6$mT). It is most likely that the low field behavior, which does not significantly vary with temperature, is dominated by nuclear magnetism and that the nuclear contribution is effectively quenched above 6~mT. Therefore the behavior of $\lambda$ in the low field regime ($B\lesssim6$~mT) is excluded from any further discussion.

The $B$-dependence of the relaxation rate can be used to determine the nature of transport of spin excitations (i.e.\ whether ballistic or diffusive) as the spin auto-correlation functions have different spectral densities in the two cases. For diffusive transport, the spectral density $f(\omega)$ has the form $f(\omega)\sim \omega^{-1/2}$, which leads to a $\lambda\propto B^{-1/2}$ power-law relation. In contrast, for  ballistic transport, $f(\omega)$ follows a logarithmic relation $f(\omega)\sim\ln(J/\omega)$, or $\lambda\propto \ln(J/B)$.

\begin{table*}
\begin{ruledtabular}
\centering
\renewcommand{\arraystretch}{1.25}
\begin{tabular}{>{\centering\arraybackslash}m{0.65in}|>{\centering\arraybackslash}m{2.1in} >{\centering\arraybackslash}m{2.1in} >{\centering\arraybackslash}m{2.1in} }
  &Model I (diffusive): $\lambda=aB^{-n}$ &Model II-a (ballistic): $\lambda=c\ln(J/B)$ &Model II-b (ballistic): $\lambda=c\ln(J/B)$\\  &$6~\mathrm{mT}\lesssim B\lesssim100~\mathrm{mT}$ &$6~\mathrm{mT}\lesssim B\lesssim450~\mathrm{mT}$ &$50~\mathrm{mT}\lesssim B\lesssim450~\mathrm{mT}$ \\\hline
 $T=$ 0.33~K &$a=0.16(3)$, $n=0.42(4)$ &$c=0.004(1)$, $J=1.4(4)$~T &$c=0.0025(2)$ \\\hline
 $T=$ 1.40~K &$a=0.17(4)$, $n=0.42(5)$ &$c=0.005(1)$, $J=1.1(3)$~T &$c=0.0026(2)$\\\hline
 $T=$ 15~K &$a=0.13(2)$, $n=0.21(3)$ &$c=0.009(1)$, $J=1.8(5)$~T &$c=0.0063(3)$\\
 \end{tabular}
     \end{ruledtabular}
\caption{Fitted coefficients 
for Cu(pyz)(NO$_3$)$_2$. The exchange $J$ was fixed at the known value of 7.2~T in Model III.}\label{tab:fit-results}
\end{table*}

Alternative fits based on two models were used to describe the relaxation rate (see Table~\ref{tab:fit-results}). In the temperature regime $T_{\mathrm{N}}<T<J $, where we expect the spin transport models to apply, 
we first attempt to fit all data for $B\gtrsim6$~mT to the ballistic transport function $\lambda=c\ln(J/B)$ (Model II-a).
This results in unacceptably small values of the intrachain exchange constant $J$, where we obtain $J=1.4(4)$~T at $T=$ 0.33~K and $1.1(3)$~T at 1.40~K.  These compare poorly with the value $J=7.2$~T, inferred from magnetic susceptibility measurements and suggests that the ballistic model does not apply over this range. We also tried to fit all data for $B\gtrsim6$~mT to a power law function $\lambda=aB^{-n}$ with a constant $a$, but the fit quality is poor due to the fact that $\lambda$ flattens out above 100 mT at both $T=0.33$~K and $T=1.40$~K. 

We find that the lower field region of the data can be best described with the power law expression (Model I) for $B$ in the range $6~\mathrm{mT}\lesssim B\lesssim100~\mathrm{mT}$ (before $\lambda$ flattens out at higher fields) [solid lines in Fig.~\ref{fig:fit-parameters}(a), (b)]. Fitting this model to the measured data gives $n= 0.42(4)$ at  $T= 0.33$~K and  $n=0.42(5)$ at 1.40~K.  It is worth noting that although recent numerical calculation~\cite{znidaric-prl-2011} suggests that spin transport in the isotropic Heisenberg chain is anomalous with $n=0.25$ rather than 0.5, our results are better matched to the theoretical prediction $n=0.5$ given by the classical model. 


Above around 100~mT the data is no longer well described by Model I. Intriguingly, we find that for $B\gtrsim50$~mT, where the power law fit fails,  the data can be fitted to the predictions of the ballistic model reasonably well in the limit of high fields, with $J$ fixed at the experimentally determined value of 7.2~T (Model II-b). This raises the possibility that the muon probe response crosses over from detecting diffusive behavior at low fields (i.e.\ at low frequency) to ballistic behavior (on a shorter time scale) above $B\approx 50$~mT. This might be expected in that diffusion seen at short times (or high frequency) should start with ballistic steps. In the present case it is possible that we are able to detect both of these parts of the fluctuation spectrum by tuning the frequency response window of the muon probe with applied magnetic field.  However, there is reason to be cautious here owing to the significant experimental uncertainty arising from the sensitivity of the results to detector dead times and the resolution limit of the spectrometer for the small relaxation rates measured for fields $B\gtrsim 200$~mT.

We may compare these results with those found outside the region of applicability of the spin transport models. 
In the high temperature regime $T>J$ the failure of both models is expected since the muons are responding not only to delocalised spin excitations but also to the quasi-independent spin flips introduced by thermal fluctuations. 
As expected, fits of our measurements made at $T=15$~K [Fig.~\ref{fig:fit-parameters}(c) and  Table~\ref{tab:fit-results}] show that neither the diffusive model nor the ballistic models return physically realistic parameters. 

We may also compare our results with that found in DEOCC-TCNQF$_{4}$, where diffusive transport was also reported. We see that our measurements of diffusion are in better agreement with the standard diffusion theory ($n\approx0.42$ for Cu(pyz)(NO$_3$)$_2$, as opposed to $n\approx 0.35$ reported for DEOCC-TCNQF$_{4}$ in Ref~\onlinecite{francis-prl-2006}) even though the latter is predicted to be a better isolated 1DQHAF on the basis of its lack of magnetic ordering down to 20~mK. In DEOCC-TCNQF$_{4}$ the diffusive behavior shows a low-field cut-off, from which the interchain-intrachain ratio $J_\perp/J_\parallel$ can be estimated.  
This cut-off should scale with $J_{\perp}$, and should therefore occur in  Cu(pyz)(NO$_3$)$_2$ at $B\approx2$~mT. However, no sharp change in behavior is observed at this low field; this is most likely due to the electronic contribution to the relaxation being swamped by the nuclear contribution at this low field.

The comparison with  DEOCC-TCNQF$_{4}$, also allows a further consistency check with the analysis of the muon site based on our DFT calculations. By comparing the 
relaxation rates found at 10~mT in the two materials, we are able to estimate the hyperfine coupling via the scaling relation \cite{francis-prl-2006} $(A_{1}/A_{2})^2=(J_{1}/J_{2})^{1/2}(\lambda_{1}/\lambda_{2})$. This approach predicts a hyperfine coupling in Cu(pyz)(NO$_3$)$_2$ of 37~MHz, in very good agreement with the first-principles value of 38 MHz found for the nitrate sites shown in Fig~\ref{fig:structures}(a) (see Table~\ref{tab:hpc} in the Appendix). This lends further weight to our conclusion that it is these NO$_{3}^{-}$ muon sites that are sensitive to the intrinsic magnetism in this material and, in particular, it is this particular nitrate site that is probing the dynamics of the system in the measurements. 

To compare the behavior of the 1D chain more generally, we also made measurements of the 2D material Cu(pyz)$_2$(ClO$_4$)$_2$ in the region $T_{\mathrm{N}}< T < J$.
Similarly to the 1D chain, the asymmetry spectra of the 2D material show Kubo-Toyabe relaxation at low magnetic field which is suppressed at high fields.  A  difference observed in the asymmetry spectra between Cu(pyz)(NO$_3$)$_2$ and Cu(pyz)$_2$(ClO$_4$)$_2$ is that in addition to the slow relaxation, a fast exponential decay ($\Lambda\approx 2$~MHz) persists in the asymmetry spectra up to the highest field. The total asymmetry signal was fitted to a sum of two exponential decaying components with a background contribution:
$A(t) = A_{\mathrm{s}}G_z^\mathrm{KT}(\Delta,B){\rm e}^{-\lambda t}+A_{\mathrm{f}}{\rm e}^{-\Lambda t}+A_{\mathrm{bg}}$,
where $A_{\mathrm{s}}$ and $A_{\mathrm{f}}$ correspond to the amplitude of the Kubo-Toyabe and the fast relaxing component, respectively. Here $\lambda$ is the relaxation rate for the slow-relaxing component, $\Lambda$ is fixed at 1.65~MHz for the fast relaxing component and $A_{\mathrm{bg}}$ accounts for the background contribution.
From Fig~\ref{fig:fit-parameters}(d), it can be seen that the relaxation rate $\lambda$ shows little field dependence,  remaining constant within the experimental uncertainty. In fact, it is probable that the spectrometer is close to its rate resolution limit below 6~kHz. 

\section{Conclusion}

We have determined candidate muon stopping sites in the 1DQHAF Cu(pyz)(NO$_3$)$_2$  using DFT calculation and 
propose that a site involving the muon forming a hydroxyl bond to an oxygen on the nitrate group is sensitive to the intrinsic dynamics of propagating spin excitations in the system. Our LF $\mu^+$SR measurements of the dynamics of the 1DQHAF Cu(pyz)(NO$_{3}$)$_{2}$ show that in the temperature range $T_{\mathrm{N}}<T<J$ the transport of spin excitations detected by the muon is diffusive over much of the range of applied fields.   The results also show a possible crossover in the probe behavior in the $50~\mathrm{mT} \lesssim B \lesssim 100$~mT region, with the power law not describing the data well above these fields, suggesting that we are sensitive to the presence of an early-time ballistic regime.

\section{Acknowledgements}
We are grateful to the STFC ISIS Facility for the provision of muon beamtime, the E-Infrastructure South Initiative for CPU time and EPSRC (UK) for financial support. FX acknowledges funding from the John Templeton Foundation. We thank Daniel Khomskii for useful discussions.  The work at EWU was supported by the U.S.\ National Science Foundation under grant no. DMR-1306158. 

\appendix*
\section{Computational details}
For our DFT calculations the wavefunction and charge-density cutoffs were 80 and 320 Rydberg, respectively. Brillouin-zone integration was performed at the $\Gamma$ point.
The muon was placed in several randomly chosen sites (all with a multiplicity of 8) which were a maximum of 1.5~\AA\ from the nearest neighbor starting site (or its crystallographic equivalent). All atoms were then allowed to relax. The starting points for the structural relaxations were based on the structures measured at 2~K~\cite{jornet-somoza-ic-2010} and used the experimental unit cell parameters. All relaxations were done in a collinear spin-polarized calculation with magnetic propagation vector $(1/2~0~0)$ (for the conventional unit cell). 
Allowing the ionic positions in the supercell without the muon to relax from the experimentally known positions yields negligible relaxation ($<0.1$~\AA). For the relaxed supercell, the L\"owdin analysis yields a Cu$^{2+}$ moment of 0.56$\mu_{\rm B}$ while the integrated absolute magnetisation of one supercell is 9.1$\mu_{\rm B}$ (containing 8 Cu$^{2+}$ ions). Given the difficulties in calculating absolute magnetic moments accurately by projecting onto atomic orbitals (such as in a L\"owdin analysis), we conclude that the DFT predicts a full Cu$^{2+}$ moment of 1$\mu_{\rm B}$ in the unperturbed system. The results discussed in this section are summarized for the paramagnetic and diamagnetic muon sites in Tables~\ref{tab:hpc} and \ref{tab:sum} respectively.

\begin{figure}[ht!]
\includegraphics[width=0.68\linewidth]{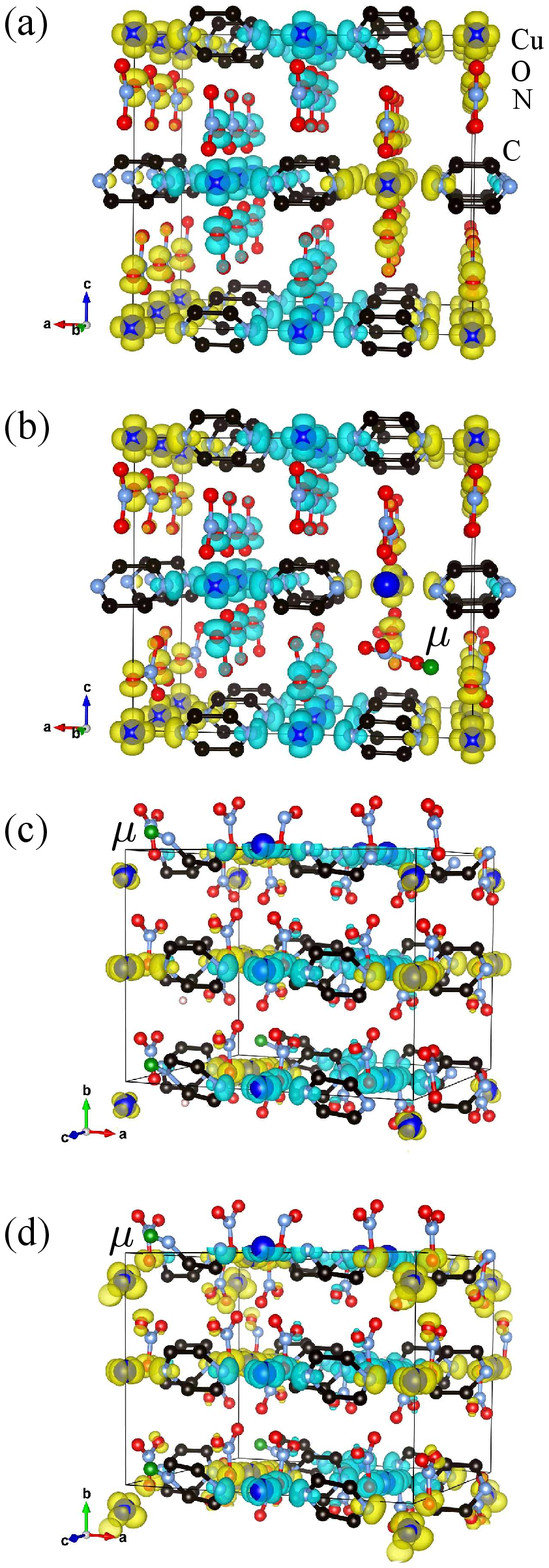}
\caption{\label{fig:spins}(color online). Calculated spin density isosurfaces drawn at $0.004/a_{0}^3$ ($a_0$ is the Bohr radius). Yellow shading is positive, blue is negative. (a) Unperturbed Cu(pyz)(NO$_3$)$_2$ without a muon showing the quasi one-dimensional exchange via the pyz rings along the $a$-axis. (b) The nitrate site with a rotated nitrate group, performed for a neutral supercell. The muon's electron has been donated to the nearest-neighbor Cu ions, turning it into diamagnetic Cu$^{+}$. (c) The N(pyz) site for a neutral supercell. The nearest-neighbor Cu ion is displaced considerably and additionally, its moment is destroyed through the donation of the muon's electron. (d) The same N(pyz) site but for a charged supercell. The Cu ion is displaced by a similar amount but it retains a magnetic moment, leading to some magnetic overlap between neighbouring Cu--pyz--Cu chains via a nitrate group.}
\end{figure}

The calculated spin density was used to assess the degree of perturbation caused by the muon probe in each of the sites identified and discussed in section II. Examples are shown for four sample sites in Fig.~\ref{fig:spins}, where  it is compared to the unperturbed structure [Fig~\ref{fig:spins}(a)]. Fig.~\ref{fig:spins}(b) shows the interruption of the magnetic exchange pathway for the nitrate site through the destruction of the Cu moment found for neutral supercells. Fig.~\ref{fig:spins}(c) shows that for the N(pyz) site in the neutral cell, there is a combined effect of (i) switching off the Cu moment and (ii) significantly displacing the Cu ion. Fig.~\ref{fig:spins}(d) shows the same scenario for the charged cell, where only the Cu ion is displaced, leading to an increased magnetic overlap between neighbouring Cu--pyz--Cu chains. 

If there is any polarised spin density $\rho({\bf r}_\mu)$ at the muon site there is a Fermi contact interaction
\begin{equation}
A=\frac{2\mu_0}{3}\gamma_\mu\gamma_e \rho({\bf r}_\mu),
\label{eqn:contact_coupl}
\end{equation}
where $\gamma_\mu$ is the muon gyromagnetic ratio and $\gamma_e$ is the electron gyromagnetic ratio. The contact hyperfine couplings were calculated using the projector augmented wave (PAW) method~\cite{Blochl1994} as implemented in the GIPAW package~\cite{qespresso} and used norm-conserving data sets with wavefunction cutoff of 120~Ry and results are given in Table~\ref{tab:hpc}. These data sets were also used in the calculation of the L\"{o}wdin charges used to estimate local magnetic moments. 

\begin{table}[htbp] 
\begin{ruledtabular}
\begin{tabular}{c l | l l l} %
 & Site & $E$ (meV) &  $A$ (MHz) \\ \hline
\multirow{2}{*}{$q=+1$} & {\bf NO$_3$} &  {\bf 110} &  {\bf 38} \\
& C(pyz) & 1200 & 370 \\ \hline
\multirow{2}{*}{$q=0$} 
& C(pyz) & 420 &  480 \\
& Interst.\ Mu & 2200 & 4640 \\
\end{tabular}
\end{ruledtabular}
\caption{\label{tab:hpc} Summary of muon sites where the contact hyperfine coupling is dominant for charged (top) and neutral supercells (bottom). Shown are the energy $E$ relative to the lowest energy site for each charge state and the contact hyperfine coupling $A$. The site probing the spin transport is highlighted in bold [shown in Fig~\ref{fig:structures}(a)].}
\end{table}

\begin{table}[htbp] 
\begin{ruledtabular}
\begin{tabular}{c l | l c} %
 & Site & $E$ (meV) & $\nu_{\rm dip}/\mu_{\rm Cu}$ (MHz$/\mu_{\rm B}$)  \\ \hline
\multirow{2}{*}{$q=+1$} & NO$_3$ &  0--200 & 1.5--12  \\
& N(pyz) & 140 & 5.5--16  \\ \hline
\multirow{3}{*}{$q=0$} & NO$_3$ (not rot.) &  0--170 & 1.1--9.6  \\
& NO$_3$ (rot.) &  70 & 1.5--9  \\
& N(pyz) & 120 & 0.4-5.6  \\
\end{tabular}
\end{ruledtabular}
\caption{\label{tab:sum} Summary of muon sites where the dipole coupling is dominant for charged (top) and neutral supercells (bottom). Shown are the energy $E$ relative to the lowest energy site for each charge state and the dipole coupling $\nu_{\rm dip}$ for the diamagnetic sites per Cu$^{2+}$ moment of 1$\mu_{\rm B}$ taking all perturbations into account.}
\end{table}

We may check the muon sites proposed above for consistency with the previous measurements\cite{tom-prb-2006} by comparing the precession frequency with that expected from the local dipole field at the muon site resulting from the long-range magnetic order. Even though the DFT calculation is unable to determine the ordered magnetic structure explicitly, the calculated supercell structural information can still be used to evaluate the dipole coupling strength at the proposed muon sites for trial structures. The dipole coupling of the muon with localised magnetic moments ${\bf m}_i$ located at position ${\bf r}_i$ is given by 
\begin{equation}
{\bf B}_{\rm dip}({\bf r}_{\mu})=\frac{\mu_0}{4\pi} \sum_i \frac{3({\bf m}_i\cdot\hat{\bf r}_{i\mu})\hat{\bf r}_{i\mu}-{\bf m}_i}{|{\bf r}_\mu - {\bf r}_i|^3},
\label{eqn:dip_field}
\end{equation}
where $\mu_0$ is the vacuum permeability, $\hat{\bf r}_{i\mu}$ the normalised vector between the muon and the moment ${\bf m}_i$. The dipolar interaction may be evaluated for an infinite sample by calculating the magnetic field given by Eq.~\ref{eqn:dip_field} within a Lorentz sphere of finite radius $r_{\rm L}$. The Lorentz sphere needs to be sufficiently large to reach satisfactory convergence of the calculated field. In our calculations $r_{\rm L}=147$~\AA. As only antiferromagnetic structures were considered, no further terms are relevant for the diamagnetic muon sites. 
Magnetic couplings of the muon and the magnetic system $\nu_{\rm dip} = \gamma_{\mu} |{\bf B}_{\rm dip}({\bf r}_{\mu})|/2\pi$ were calculated by taking into account the crystallographic distortions within one supercell around the muon (with the muon at the centre) and by taking into account perturbations to the magnetic moments, which were significant for the neutral supercell, by scaling the magnetic moment by the ratio of perturbed and unperturbed L\"owdin charge polarization on the Cu ions. 

Dipole field calculations were made for muon sites to both neutral and charged supercells for several magnetic propagation vectors. 
The following magnetic structures were considered (all for the conventional unit cell): (i) Magnetic propagation vector ($1/2$ 0 0) and the two Cu spins parallel or antiparallel, moments along any of the three crystallographic directions. (ii) Magnetic propagation vector ($1/2$ $1/2$ 0) and the two Cu spins parallel or antiparallel, moments along any of the three crystallographic directions. (iii) Magnetic propagation vector ($1/2$ $1/2$ $1/2$) and the two Cu spins parallel or antiparallel, moments along any of the three crystallographic directions. (iv) Magnetic propagation vector ($1/2$ 0 $1/2$) and the two Cu spins parallel or antiparallel, moments along any of the three crystallographic directions. (v) A spiral structure where the two Cu moments are offset by 90$^\circ$ with magnetic propagation vector ($1/2$ 0 0) with moments in the $ab$-plane. The latter structure is interesting as it represents the classical solution for a zig-zag spin ladder with frustrating next-nearest-neighbor interactions~\cite{White1996} and has recently been suggested to occur in Cu(pyz)(NO$_3$)$_2$ on the basis of electron spin resonance measurements~\cite{validov-jpcm-2014}. The calculated results are summarized in Table~\ref{tab:sum}. 
We find that couplings consistent with the experimentally observed values can be reproduced at several of the proposed sites for each of the tested magnetic structures.
It is therefore impossible to use this method to give an unambiguous assignment of the muon sites giving rise to the precession signal in the zero field $\mu^{+}$SR measurements. However, we may conclude that the nitrate sites proposed are consistent with this analysis. 


There is some evidence that the moment in Cu(pyz)(NO$_3$)$_2$ must be renormalised from its full value of 1$\mu_{\rm B}$. If the Cu moment were close to 1$\mu_{\rm B}$, we would expect to see oscillations in the range 0.4--16 MHz if we assume all low-energy sites are populated. As the observed oscillation frequencies were $\nu_1(0)=1.922(4)$~MHz and $\nu_2(0)=1.257(3)$~MHz and as very slow oscillations $\ll 0.5$~MHz would not be distinguishable from the background due to muons relaxing in the cryostat tail or sample holder, we conclude that a consistent Cu moment size would be around $\mu_{\rm Cu} \approx 0.12\mu_{\rm B}$ [$\nu_{1}(0)/16$~MHz $\mu_{\rm B}$]. The ordered moment size in low-dimensional magnetic systems is expected to be heavily renormalised by enhanced quantum fluctuations. This value is consistent with the estimate of $\mu_{\rm Cu}\approx 2.034 \sqrt{J_\perp/J}=0.136\mu_{\rm B}$ obtained from a mean field model of weakly coupled antiferromagnetic spin chains~\cite{schulz-prl-1996}, where $J=10.3(1)$~K~\cite{hammar-prb-1999} is the primary exchange and $J_\perp\approx0.046$~K is the interchain coupling~\cite{tom-prb-2006}.

Finally, in addition to the low-energy sites discussed in the main text, a high-energy radical site was found in both neutral and charged supercells, where the muon bonds to one of the four equivalent carbon atoms in the pyrazine ring. In the neutral cell, interstitial muonium (Mu) was also predicted, though at an energy considerably above the low-energy states described in the main text.  It is also conceivable that the muon may substitute for any of the four equivalent hydrogen atoms in either pyrazine ring though the energy of such a state depends on the final state of the substituted proton.

\bibliography{../../../Reference/database}

\end{document}